\RequirePackage[T1]{fontenc}
\documentclass[letterpaper,journal,10pt]{IEEEtran}

\usepackage{amsmath,amsfonts}
\usepackage{algorithmic}
\usepackage{algorithm}
\usepackage{array}
\usepackage[caption=false,font=normalsize,labelfont=sf,textfont=sf]{subfig}
\usepackage{textcomp}
\usepackage{stfloats}
\usepackage{url}
\usepackage{verbatim}
\usepackage{graphicx}
\usepackage{cite}
\usepackage{booktabs}

\begin{document}

\title{AutoC2V: Synthesis-Guided LLM Training for QoR-Optimized C++-to-Verilog Generation}

\author{Guanqi~Lyu,
Fei~Zhao,
Zhaojun~Lu,
Jiaji~He,
and Yongqiang~Lyu%

\thanks{This work has been submitted to the IEEE for possible publication.
Copyright may be transferred without notice, after which this version may
no longer be accessible.}%
\thanks{This work was supported by the National Natural
Science Foundation of China under Grant 62672345.}
\thanks{Guanqi Lyu, Zhaojun Lu, and Jiaji He
are with the School of Cybersecurity,
Tianjin University, Tianjin 300350, China
(e-mail: lyuguanqi@tju.edu.cn;
lzj\_25@tju.edu.cn;
dochejj@tju.edu.cn).}%
\thanks{Fei Zhao is with Peking University,
Beijing 100871, China
(e-mail: feizhao@pku.edu.cn).}%
\thanks{Yongqiang Lyu is with the School of
Cybersecurity, Tianjin University,
Tianjin 300350, China, and also with
Tsinghua University, Beijing 100084, China
(e-mail: lyuyq@tju.edu.cn).}%
\thanks{Corresponding author: Fei Zhao
(e-mail: feizhao@pku.edu.cn).}%
}

\maketitle

\begin{abstract}
High-level synthesis (HLS) can translate C/C++ kernels into register-transfer level (RTL) designs, yet generating high-quality hardware still relies heavily on manual pragma tuning and repeated synthesis evaluation. We propose AutoC2V, a C++-to-Verilog generation framework driven by real synthesis feedback that enables large language models (LLMs) to directly generate RTL with improved quality of results (QoR) from synthesizable C++ kernels. We construct training data from real synthesis results, including synthesizable C++ source code, the corresponding Verilog RTL, QoR metrics, and optimization objectives. Using this dataset, we align the model toward synthesis-aware RTL generation through Supervised Fine-Tuning (SFT) and Group Relative Policy Optimization (GRPO). To reduce the high cost of repeatedly invoking real synthesis during GRPO, we design the QoR Prediction Reward Model (QoR RM), which predicts latency and resource metrics by encoding RTL into long-context semantic representations, thereby providing low-cost and continuous QoR reward signals for policy optimization. Experimental results demonstrate that AutoC2V not only substantially improves RTL functional pass rates, but also achieves superior latency and resource optimization compared to the unfine-tuned base model, representative HLS optimization approaches (HLSPilot, LLM-DSE, and AutoDSE), and advanced general-purpose commercial models (Claude Opus 4.8, GPT-5.5, Gemini 3.1 Pro, DeepSeek v4 Pro, and GLM-5.2). These results indicate that combining real synthesis data with efficient QoR reward modeling can further improve the ability of LLMs to generate high-quality RTL.
\end{abstract}


\section{Introduction}
\label{sec:introduction}

\IEEEPARstart{H}{igh-level} synthesis (HLS) has become an important methodology for improving hardware design productivity by translating algorithmic descriptions in C/C++ into register-transfer level (RTL) implementations \cite{cong2011hls}. By abstracting away low-level implementation details, HLS enables designers to explore alternative hardware architectures without manually developing complete RTL designs. Nevertheless, the quality of results (QoR) of HLS-generated
hardware remains highly sensitive to synthesis directives,
commonly known as pragmas, which govern loop scheduling,
computational parallelism, and memory organization---key
factors affecting hardware latency and resource utilization
~\cite{sohrabizadeh2022autodse,amd2023vitis}. Different pragma configurations applied to the same C++ kernel can lead to substantially different latency and resource utilization. Consequently, obtaining high-quality hardware implementations often requires extensive design-space exploration (DSE) and repeated synthesis evaluations.

To reduce the effort associated with HLS optimization, numerous approaches have been proposed for automated pragma tuning and QoR prediction. AutoDSE \cite{sohrabizadeh2022autodse} employs bottleneck-guided optimization to explore HLS design spaces, while HARP \cite{sohrabizadeh2023harp} leverages hierarchical graph representations and graph neural networks to support QoR prediction and design exploration. More recently, large language models (LLMs) have been incorporated into HLS optimization workflows. HLSPilot \cite{xiong2024hlspilot} integrates LLM-assisted code transformation with pragma optimization, whereas LLM-DSE \cite{wang2025llmdse} employs multiple LLM agents to search for effective accelerator configurations. These approaches have demonstrated the potential of automated hardware optimization. Despite these advances, these approaches primarily search over
source-level transformations and HLS configurations, while the
resulting RTL architectures are ultimately determined by
conventional HLS compilers. The synthesis feedback obtained
in these workflows guides high-level design exploration rather
than directly training a QoR-aware RTL generation policy.
This suggests a complementary direction: leveraging
synthesis-generated RTL implementations and their associated
QoR measurements as supervision for training a generative model
that directly produces objective-conditioned RTL without
invoking HLS at inference time.

In parallel with HLS optimization, recent advances in
LLM-based RTL generation and QoR prediction provide a
foundation for direct, objective-conditioned hardware
generation. VerilogEval \cite{liu2023verilogeval} and RTLLM \cite{lu2024rtllm} have established benchmarks for evaluating Verilog generation and functional correctness. Beyond natural-language-driven RTL generation, Liao et al. \cite{liao2024llmhls} investigated the direct translation of C programs into Verilog using LLMs and compared the resulting hardware implementations with conventional HLS outputs. Furthermore, BetterV \cite{pei2024betterv} demonstrated that task-specific discriminative guidance can improve Verilog generation for downstream electronic design automation (EDA) objectives, while DeepRTL2 \cite{liu2025deeprtl2} showed that learned RTL representations can support hardware performance prediction. These studies provide complementary foundations for direct
high-level-to-RTL translation, objective-guided RTL generation,
and predictive QoR evaluation. Building on these capabilities,
we investigate whether RTL implementations and QoR annotations
obtained from real synthesis can be used to train a generation
policy that preserves the functionality of the input C++
program while optimizing latency or resource usage according
to the specified objective.

Realizing this goal requires overcoming three challenges. First, effective QoR-aware RTL generation requires training data that associate high-level computations with hardware implementations and their synthesis-derived QoR. Unlike conventional code-generation datasets, such data must capture the relationships among C++ semantics, synthesis directives, RTL structures, and latency--resource trade-offs. Second, optimizing generated RTL must be subject to functional correctness constraints. Improvements in hardware efficiency are meaningful only when the resulting implementations preserve the behavior of the original C++ kernels. Although supervised fine-tuning (SFT) can establish fundamental RTL generation capabilities, its token-level objective does not directly optimize functional validity or hardware quality. Third, directly incorporating synthesis feedback into reinforcement learning incurs substantial computational overhead, as policy optimization requires evaluating numerous generated RTL candidates. An effective training framework therefore requires a computationally efficient approximation of synthesis-derived QoR to provide continuous optimization signals.

To address these challenges, we propose AutoC2V, a synthesis-guided LLM training framework for objective-conditioned C++-to-Verilog generation. Given a synthesizable C++ kernel and either a latency-oriented or resource-oriented optimization objective, AutoC2V directly generates a complete Verilog RTL implementation without invoking HLS during inference. The framework acquires hardware optimization knowledge from an offline collection of synthesis-generated design points. Specifically, we construct a QoR-annotated dataset through C++ kernel preprocessing, region-level decomposition, pragma-space exploration, and real synthesis evaluation. The resulting dataset comprises 34,078 design samples derived from 201 independently synthesizable regions. Each sample associates the source C++ computation and its synthesis configuration with the corresponding Verilog RTL implementation, latency, and FPGA resource metrics. By retaining hardware implementations spanning diverse QoR levels, the dataset provides supervision for learning hardware design decisions under different optimization objectives rather than merely reproducing a single fixed synthesis result.

\begin{figure*}[!t]
\centering
\includegraphics[width=\textwidth]
{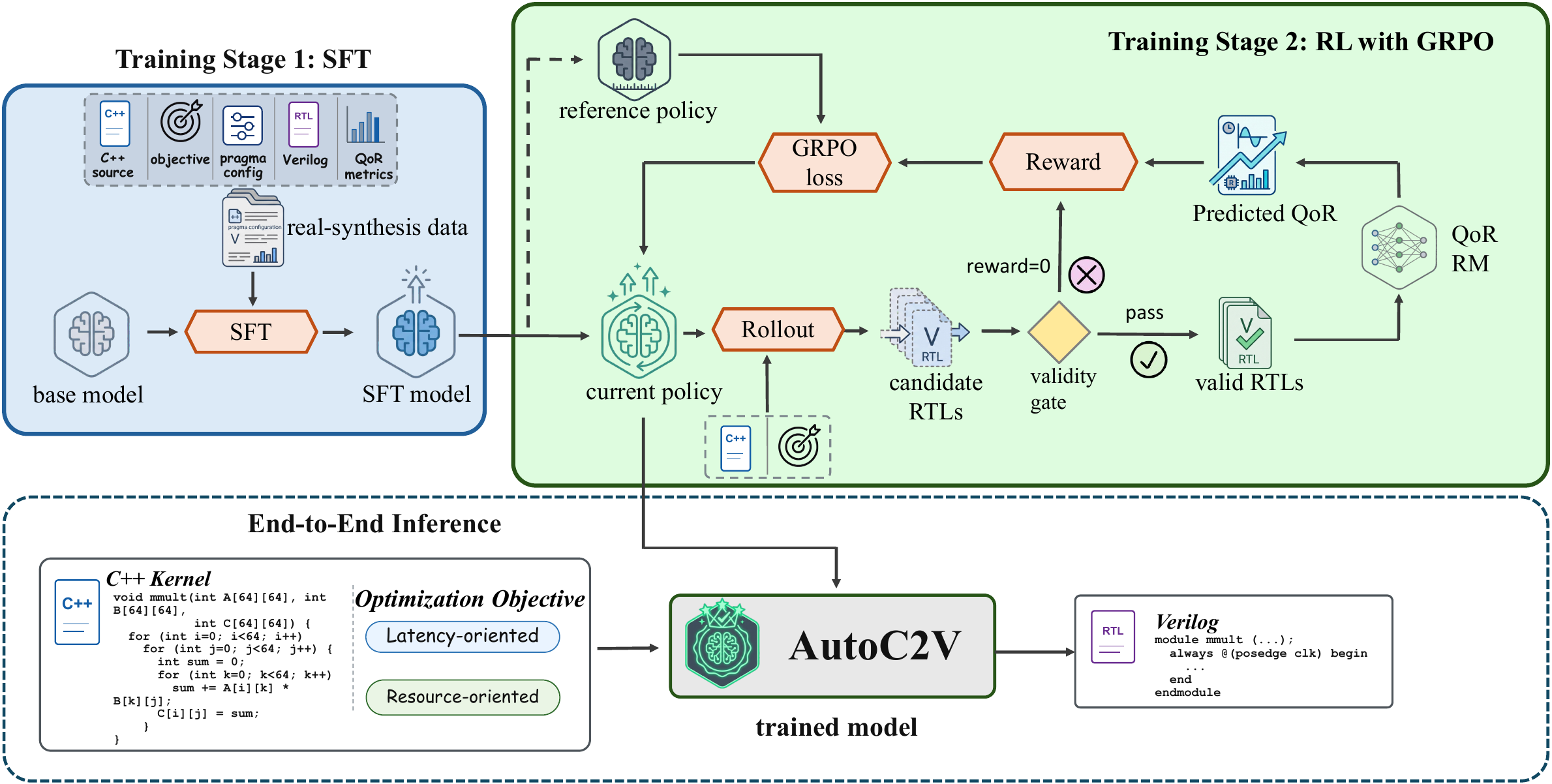}
\caption{Overview of the AutoC2V workflow.}
\label{fig:autoc2v_workflow}
\end{figure*}

AutoC2V employs a two-stage training strategy to transform synthesis-derived knowledge into an RTL generation policy. In the first stage, SFT trains the model to learn the mapping from synthesizable C++ kernels and optimization objectives to complete Verilog RTL implementations. In the second stage, Group Relative Policy Optimization (GRPO) \cite{shao2024deepseekmath} further refines the generation policy using functionality-gated, synthesis-informed rewards. Candidate RTL implementations that fail the training-time functional checks receive zero reward, whereas passing candidates are scored according to their predicted QoR. To reduce the cost of reward evaluation, we employ a QoR Prediction Reward Model (QoR RM) that combines long-context RTL representations obtained from Dewey \cite{zhang2025dewey} with XGBoost regressors \cite{chen2016xgboost}. The reward model predicts synthesis-derived changes relative to baseline designs in terms of latency and FPGA resource utilization, including DSP, BRAM, LUT, and FF usage. These predictions are subsequently converted into objective-dependent reward signals. In this way, synthesis knowledge acquired through offline hardware evaluation can guide policy optimization without requiring full synthesis for every generated candidate. Fig.~\ref{fig:autoc2v_workflow} shows the overall AutoC2V workflow.

We evaluate AutoC2V on 59 held-out kernels derived from HLS-Eval \cite{abikaram2025hlseval}, assessing RTL synthesizability, functional correctness, and hardware QoR. Experimental results show that AutoC2V achieves a single-candidate synthesis pass rate (Syn@1) of 80.5\% and a functional pass rate (Func@1) of 66.9\%, averaged over the two optimization objectives, substantially outperforming the SFT-only baseline in functional correctness. AutoC2V also achieves superior latency and resource optimization compared with the unfine-tuned base model, representative HLS optimization approaches (HLSPilot, LLM-DSE, and AutoDSE), and general-purpose commercial LLMs (Claude Opus 4.8, GPT-5.5, Gemini 3.1 Pro, DeepSeek v4 Pro, and GLM-5.2). These findings demonstrate the potential of combining real synthesis knowledge with reinforcement learning to generate functionally correct RTL implementations with favorable hardware QoR.

The main contributions of this work are summarized as follows:

\begin{itemize}

\item \textbf{Synthesis-supervised C++-to-RTL generation.}
We develop an objective-conditioned generation framework that learns hardware optimization knowledge from real synthesis-generated design points and directly produces complete Verilog RTL implementations from synthesizable C++ kernels. We construct a synthesis-annotated training dataset containing diverse RTL implementations and their latency and resource characteristics, enabling the model to learn hardware design decisions under different QoR objectives.

\item \textbf{Functionality-gated QoR policy optimization.}
We integrate SFT with GRPO to optimize RTL generation under functional correctness constraints and objective-dependent hardware quality preferences. A synthesis-trained QoR prediction model provides continuous reward signals for latency and FPGA resource optimization, enabling reinforcement learning without invoking full synthesis for every generated candidate.

\item \textbf{Experimental evaluation.}
We evaluate AutoC2V on held-out HLS kernels using synthesizability checks, functional verification, and hardware QoR measurements. Experimental results demonstrate improvements in RTL functional correctness and objective-conditioned QoR compared with the SFT-only model.

\end{itemize}

\section{Background and Motivation}
\label{sec:background}

\subsection{HLS Compilation and QoR Optimization}
\label{subsec:hls_background}

HLS implements a synthesizable C/C++ computation through operation scheduling, resource allocation and binding, and control generation~\cite{cong2011hls}. Scheduling determines when operations execute, while allocation and binding determine which operators and storage elements implement them. Consequently, the same source computation can yield different datapaths, schedules, and memory organizations under different directives and target constraints.

Three directive families define the design spaces considered in this work~\cite{amd2023vitis}. \texttt{PIPELINE} overlaps successive loop iterations; its initiation interval (II) specifies the cycles between their starts. The requested II may be limited by dependencies, operator availability, or memory ports. \texttt{UNROLL} exposes iteration-level parallelism by replicating loop bodies. \texttt{ARRAY\_PARTITION} increases available memory ports by dividing an array into banks or individual elements. Their effects are coupled: additional arithmetic lanes require sufficient operand bandwidth, and greater parallelism can increase resource usage or complicate timing closure. 

We characterize QoR by transaction latency and four FPGA resource metrics: digital signal processing blocks (DSPs), block RAM (BRAM), lookup tables (LUTs), and flip-flops (FFs). Transaction latency measures the completion time of a kernel invocation. Resource metrics capture distinct implementation costs: operator sharing can reduce DSP usage, memory banking can change BRAM and LUT usage, and pipelining can increase FF usage. These quantities therefore describe a latency--resource trade-off rather than a single measure of hardware quality.

For the final hardware evaluation in this work, latency is measured in RTL simulation from the first sampled assertion of the start request to completion, including any wait for request acceptance. Resources are obtained from Vivado out-of-context RTL synthesis under a common device and clock constraint.

\subsection{C++-to-RTL Generation}
\label{subsec:generation_paradigms}

Conventional HLS flows optimize C/C++ source code and synthesis directives, then use an HLS compiler to generate RTL. Design-space exploration automates the selection of these high-level design choices, often aided by learned QoR predictors. LLM-assisted HLS uses language models to transform source code or suggest directives, while the HLS compiler still produces the RTL. Alternatively, an LLM can translate the input program directly into RTL~\cite{liao2024llmhls}, generating the datapath, control logic, and interfaces as Verilog code.

AutoC2V adopts direct RTL generation and learns from offline HLS design points. Given a synthesizable C++ kernel $x$ and an optimization objective $o\in\{\mathrm{latency},\mathrm{resource}\}$, the model outputs a complete Verilog hierarchy $v$:
\begin{equation}
    v\sim\pi_{\theta}(\cdot\mid x,o).
    \label{eq:background_generation}
\end{equation}

HLS compilation supplies the RTL implementations and QoR annotations during offline data construction. At inference time, AutoC2V generates Verilog directly without invoking HLS. 

\subsection{Motivating Example and Key Challenges}
\label{subsec:motivating_example}

As a conceptual illustration, consider the vector multiply-add kernel in Fig.~\ref{fig:motivating_kernel}, assuming that the arrays occupy disjoint storage. Each iteration updates a different output element, and assignment to \texttt{ap\_uint<32>} retains the result modulo $2^{32}$.

\begin{figure}[!t]
\centering
\begin{minipage}{0.98\columnwidth}
\normalsize
\begin{verbatim}
#include <ap_int.h>
void vmadd(ap_uint<32> a[64],
           ap_uint<32> b[64],
           ap_uint<32> c[64]) {
  for (int i = 0; i < 64; ++i)
    c[i] = a[i] * b[i] + c[i];
}
\end{verbatim}
\end{minipage}
\caption{Illustrative vector multiply-add kernel with 32-bit unsigned modular arithmetic.}
\label{fig:motivating_kernel}
\end{figure}

A resource-oriented design can reuse arithmetic operators across iterations. A latency-oriented design can overlap iterations or expose multiple arithmetic lanes, provided that the memory and interfaces supply the required bandwidth. Unrolling without adequate memory ports can leave the design bandwidth-limited; additional pipeline stages can improve scheduling or timing while increasing registers. Selecting an implementation therefore requires coordinating computation, storage, and control rather than independently choosing aggressive directive values.

These structural changes must preserve every output update and the completion protocol. Missing writes, incorrect truncation, or premature completion invalidate an otherwise plausible datapath. The example also exposes the feedback problem: syntax checks cannot establish either behavioral correctness or QoR, while synthesizing every candidate is expensive. AutoC2V therefore learns coordinated implementation choices from synthesis-derived examples and uses functional checks followed by predicted QoR during policy training. 
\section{Related Work}
\label{sec:related_work}

\subsection{HLS Design-Space Exploration and QoR Prediction}
\label{subsec:related_hls_dse}

Automated HLS optimization combines design-space search with hardware evaluation. AutoDSE~\cite{sohrabizadeh2022autodse} uses bottleneck-guided coordinate optimization to concentrate on high-impact parameters. GNN-DSE~\cite{sohrabizadeh2022gnndse} learns graph-based surrogates for HLS evaluation, while HARP~\cite{sohrabizadeh2023harp} introduces hierarchical representations to improve HLS modeling and exploration. IronMan-Pro~\cite{wu2023ironmanpro} integrates graph-based performance and resource prediction, reinforcement learning for multiobjective resource allocation, and code transformation into an HLS optimization framework.

These methods demonstrate that learned evaluation and reinforcement learning can reduce exploration cost and improve HLS designs. Their optimized decisions are realized through high-level compilation. 

\subsection{LLM-Assisted HLS Optimization}
\label{subsec:related_llm_hls}

HLSPilot~\cite{xiong2024hlspilot} combines application profiling, LLM-assisted C/C++-to-HLS transformation, and pragma-parameter DSE. LLM-DSE~\cite{wang2025llmdse} coordinates routing, specialist, arbitration, and critic agents to refine accelerator directives using tool feedback. Both integrate language-based reasoning with search while retaining HLS as the RTL-generation mechanism.

LIFT~\cite{prakriya2025lift} transfers structural information into LLM training through GNN-supervised pragma insertion. HLStrans~\cite{zou2025hlstrans} provides paired C and HLS programs, testbenches, and synthesis annotations for learning high-level hardware transformations. These works are closely related to learning synthesis-informed decisions from data.

\subsection{LLM-Based RTL Generation and High-Level Translation}
\label{subsec:related_rtl_generation}

VeriGen~\cite{thakur2024verigen} adapts language models to Verilog through domain-specific fine-tuning. RTLFixer~\cite{tsai2024rtlfixer} uses retrieval and compiler feedback to repair generated Verilog syntax. VerilogEval~\cite{liu2023verilogeval} evaluates generated RTL with simulation-based functional tests, and RTLLM~\cite{lu2024rtllm} considers syntax, functionality, and design quality for natural-language-driven generation. These studies establish generation and evaluation techniques relevant to AutoC2V, although their specifications differ from executable C++ inputs.

Liao et al.~\cite{liao2024llmhls} explicitly investigate direct C-to-Verilog translation and compare LLM-generated hardware with conventional HLS. Their study uses ChatGPT-4o on nine PolyBench kernels and reports repeated prompting to construct hardware projects.

\subsection{QoR-Guided RTL Generation and Predictive Evaluation}
\label{subsec:related_qor_rtl}

BetterV~\cite{pei2024betterv} combines Verilog fine-tuning with task-specific generative discriminators for controlled generation, demonstrating reductions in synthesized netlist nodes and SAT-based verification runtime. VeriOpt~\cite{tasnia2025veriopt} uses specialized LLM roles and EDA feedback for PPA-aware generation. REvolution~\cite{min2025revolution} combines LLM generation with evolutionary search, organizing candidates into separate populations for functional repair and PPA optimization. These methods establish that downstream hardware objectives can guide RTL generation through learned control or feedback-based search.

RTL representations also support predictive evaluation. DeepRTL2~\cite{liu2025deeprtl2} unifies generation and embedding tasks and predicts area and delay using learned RTL embeddings with separate XGBoost regressors. AutoC2V builds on representation-based regression, using Dewey~\cite{zhang2025dewey} and XGBoost~\cite{chen2016xgboost} to predict baseline-relative changes in latency, DSP, BRAM, LUT, and FF for objective-dependent reward computation.

\section{Proposed Methodology}
\label{sec:methodology}

\subsection{Data Construction}
\label{subsec:method_data}

To construct a high-quality training set, we start from candidate C++ samples in HLStrans~\cite{zou2025hlstrans} and apply cleaning, repair, deduplication, and synthesizability filtering. For incomplete kernels, we use Qwen3-Next-80B-A3B-Thinking~\cite{qwen2025next} to add only the missing headers, type definitions, and declarations while preserving the original structure and interfaces. We then retain unique, synthesizable kernels with distinct top-function names for downstream region decomposition and real-synthesis-guided diverse sampling.

To expose finer-grained program structures and RTL generation patterns, we use GPT-5.5 to decompose each unique kernel along complete loop nests, function boundaries, and algorithmic stages. Each region is then reorganized into a standalone region-level C++ kernel and checked for independent synthesizability, yielding 201 synthesizable regions for pragma-space construction and real-synthesis-guided diverse sampling.

For each region, we use libclang~\cite{lattner2004llvm} to extract loops, arrays, function boundaries, memory accesses, and top-function metadata. We use this structural information to construct candidate spaces for \texttt{PIPELINE}, \texttt{UNROLL}, and \texttt{ARRAY\_PARTITION}. These pragmas control loop parallelism and memory bandwidth, two major drivers of the latency--resource trade-off in HLS. A pragma site denotes a specific loop or array declaration at which a pragma can be inserted, while a state denotes a concrete setting available at that site.

We explore each pragma space through real-synthesis-guided diverse sampling, ranking synthesized designs by the target-oriented score for the current objective. Defined in the following subsection, this score summarizes latency and resource improvements relative to a baseline synthesized from the same region with all search-space pragmas removed. Single-site screening evaluates one candidate state at a single site while leaving all other sites in the no-pragma state; elite mutation perturbs a few sites in high-scoring designs; interaction expansion combines previously effective states; QoR-range repair targets underrepresented score ranges; and random exploration samples novel configurations. Each candidate is materialized as a pragma-annotated C++ design and compiled into a complete Verilog hierarchy using Vitis HLS. The resulting RTL is evaluated to obtain latency, DSP, BRAM, LUT, and FF metrics. The hardware evaluation protocol is specified in the experimental setup.

\begin{figure*}[!t]
\centering
\includegraphics[width=\textwidth]{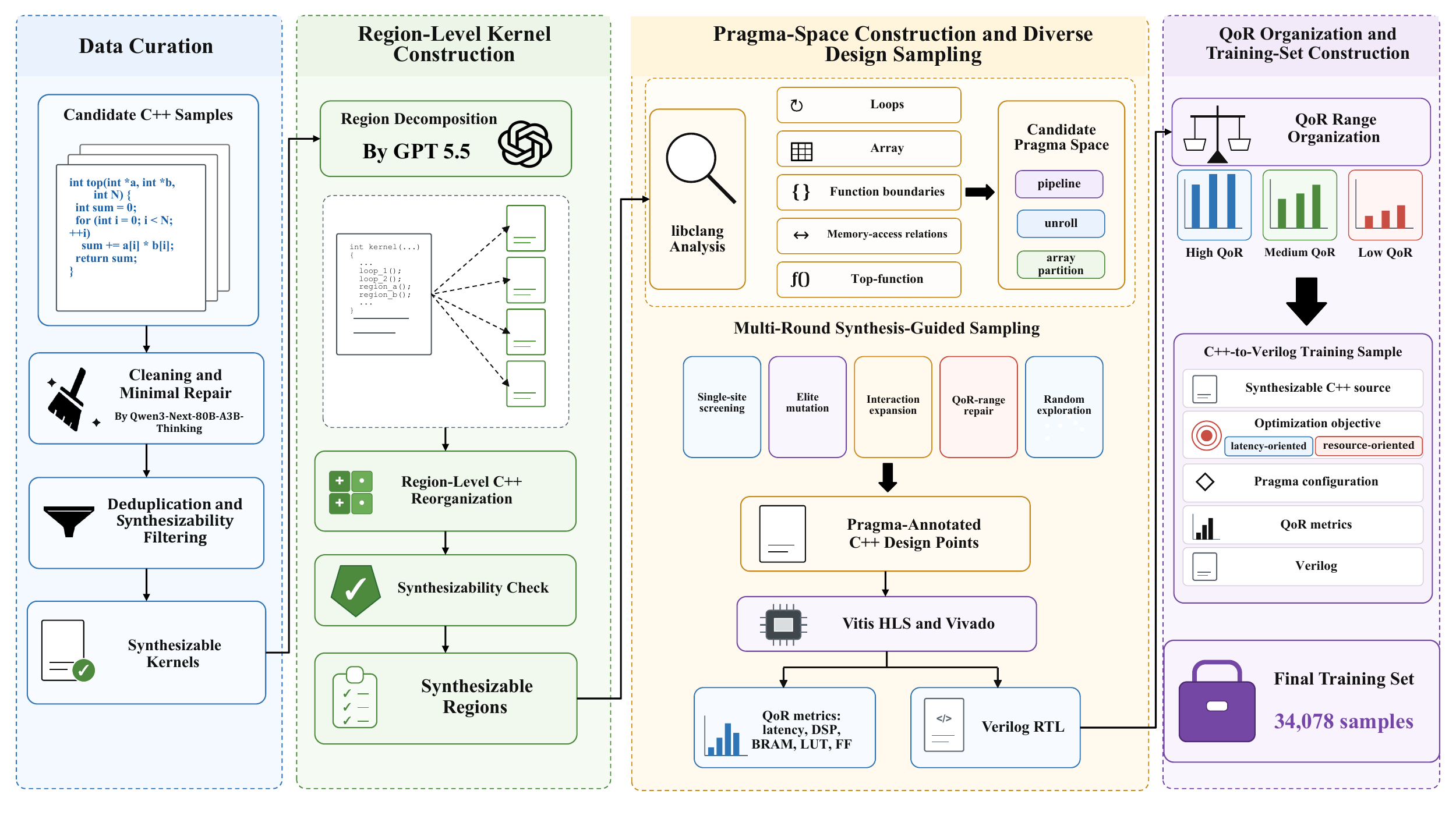}
\caption{Pipeline for data preprocessing and training-set construction.}
\label{fig:method_data}
\end{figure*}

According to the sampling objective, we retain synthesized designs across high, medium, and low ranges of the target-oriented score. Each of the resulting 34,078 samples contains synthesizable C++ source, an optimization objective, real-synthesis QoR metrics, and Verilog RTL, as shown in Fig.~\ref{fig:method_data}.

\subsection{QoR Normalization and Scoring}
\label{subsec:method_scoring}

To uniformly measure design quality across different kernels and optimization objectives, we convert real-synthesis QoR into normalized utilities relative to a corresponding baseline. For each candidate design $d$, the corresponding baseline is synthesized from the same region-level kernel after removing the three pragma types in our search space (\texttt{PIPELINE}, \texttt{UNROLL}, and \texttt{ARRAY\_PARTITION}) while leaving the source otherwise unchanged. Let the candidate QoR vector be $\mathbf q$, and let the baseline QoR vector be $\mathbf b$. For each metric $m\in\mathcal M=\{\mathrm{latency},\mathrm{DSP},\mathrm{BRAM},\mathrm{LUT},\mathrm{FF}\}$, we first compute the log ratio:
\begin{equation}
r_m=\log\frac{q_m+1}{b_m+1}.
\label{eq:method_logratio}
\end{equation}
Here, $q_m$ and $b_m$ denote the synthesis results of the candidate design and the baseline for metric $m$, respectively. The log ratio reduces QoR-scale differences across kernels, while adding one handles zero-valued resources. We clip $r_m$ to $[-c,c]$, obtaining
\begin{equation}
\widetilde r_m=\operatorname{clip}(r_m,-c,c).
\label{eq:method_clipped_ratio}
\end{equation}
In our implementation, $c=2.5$, which bounds the corresponding multiplicative ratio to $[e^{-2.5},e^{2.5}]\approx[0.082,12.2]$. This interval captures substantial optimization effects while preventing extreme synthesis results from dominating the score.

Because latency and resource usage are both minimized but generally trade off against one another, we first map their clipped relative changes to separate component utilities and then combine them according to the requested optimization objective. Let $\mathcal R=\{\mathrm{DSP},\mathrm{BRAM},\mathrm{LUT},\mathrm{FF}\}$ denote the set of resource metrics. The two component utilities are defined as
\begin{equation}
\begin{aligned}
U_{\mathrm{lat}}&=\operatorname{clip}_{[u_{\min},u_{\max}]}\bigl(\sigma(-\kappa\widetilde r_{\mathrm{latency}})\bigr),\\
\widetilde r_{\mathrm{res}}&=\sum_{m\in\mathcal R}w_m\widetilde r_m,\\
U_{\mathrm{res}}&=\operatorname{clip}_{[u_{\min},u_{\max}]}\bigl(\sigma(-\kappa\widetilde r_{\mathrm{res}})\bigr).
\end{aligned}
\label{eq:method_utilities}
\end{equation}
Here, $U_{\mathrm{lat}}$ and $U_{\mathrm{res}}$ denote the latency and resource utilities, respectively, where larger values indicate better quality in the corresponding dimension. $\sigma(t)=(1+e^{-t})^{-1}$ is the sigmoid function, $\kappa$ controls the mapping slope, and $[u_{\min},u_{\max}]$ defines the score bounds. We use $\kappa=2.0$ and $[u_{\min},u_{\max}]=[0.01,0.99]$, which allows moderate QoR improvements to be distinguished effectively while avoiding complete score saturation. When a candidate design has the same QoR as the baseline, the corresponding log ratio is zero and the utility equals $0.5$; therefore, utility values above $0.5$ indicate improvement over the baseline, while values below $0.5$ indicate degradation.

We use $w_{\mathrm{DSP}}=0.35$, $w_{\mathrm{BRAM}}=0.25$, $w_{\mathrm{LUT}}=0.25$, and $w_{\mathrm{FF}}=0.15$ as the default resource preference in this study. These user-configurable weights reflect a prior preference over FPGA resource types rather than a universal resource model. DSP receives the largest weight because dedicated arithmetic resources often constrain accelerator parallelism; BRAM and LUT receive medium weights to account for on-chip storage and logic usage; and FF receives a smaller weight because register usage can naturally increase with pipelining. The same preference is kept fixed across data construction, model training, and final evaluation.

For target-oriented evaluation, we further combine the two component utilities:
\begin{equation}
\begin{aligned}
S_{\mathrm{lat}}&=0.8U_{\mathrm{lat}}+0.2U_{\mathrm{res}},\\
S_{\mathrm{res}}&=0.2U_{\mathrm{lat}}+0.8U_{\mathrm{res}}.
\end{aligned}
\label{eq:method_objective_scores}
\end{equation}
Here, $S_{\mathrm{lat}}$ is used for latency-oriented prompts, and $S_{\mathrm{res}}$ is used for resource-oriented prompts. Similar to the resource weights, the $0.8/0.2$ objective weights are user-configurable defaults in this study. They emphasize the requested primary objective while retaining the secondary objective as a soft preference rather than a hard constraint.

$S_{\mathrm{lat}}$ and $S_{\mathrm{res}}$ are used for organizing the training archive and reporting final QoR statistics. During data construction, each target-oriented score is assigned to one of five equal-width bins over the unit interval, and samples are drawn across the populated bins to retain high-, medium-, and low-quality design examples.

\subsection{Training}
\label{subsec:method_training}

AutoC2V adopts a two-stage training pipeline consisting of SFT and GRPO. In the first stage, SFT learns a stable C++-to-Verilog output format, enabling the model to generate complete Verilog RTL from a synthesizable C++ kernel and an optimization objective. Let $\mathcal D_{\mathrm{SFT}}$ denote the empirical distribution of SFT training samples $(x,o,v^*)$, where $x$ is the input kernel, $o$ is a latency-oriented or resource-oriented objective, and $v^*$ is the target Verilog hierarchy corresponding to a real-synthesis design point. The SFT stage minimizes the autoregressive negative log-likelihood:
\begin{equation}
\mathcal L_{\mathrm{SFT}}(\theta)=-\mathbb E_{(x,o,v^*)\sim\mathcal D_{\mathrm{SFT}}}\left[\sum_{t=1}^{|v^*|}\log\pi_\theta(v_t^*\mid x,o,v_{<t}^*)\right].
\label{eq:method_sft}
\end{equation}

In the second stage, we adopt GRPO for target-oriented reinforcement learning alignment~\cite{shao2024deepseekmath}. For each input prompt $(x,o)$, the old policy $\pi_{\theta_{\mathrm{old}}}$ samples a group of candidate outputs $\{v_i\}_{i=1}^G$, where $G$ denotes the rollout group size. Each candidate is scored by the reward function $R_i=R(x,o,v_i)$, and the rewards are normalized within the sampled group to form group-relative advantages:
\begin{equation}
A_i=\frac{R_i-G^{-1}\sum_{j=1}^{G}R_j}{\operatorname{std}(R_1,\ldots,R_G)+\epsilon_{\mathrm{adv}}}.
\label{eq:method_advantage}
\end{equation}
Here, $\epsilon_{\mathrm{adv}}$ is a small constant for numerical stability.

GRPO then updates the model using a clipped policy-ratio objective. Specifically, the token probability ratio between the current policy $\pi_\theta$ and the old policy $\pi_{\theta_{\mathrm{old}}}$ is
\begin{equation}
\rho_{i,t}(\theta)=\frac{\pi_\theta(v_{i,t}\mid x,o,v_{i,<t})}{\pi_{\theta_{\mathrm{old}}}(v_{i,t}\mid x,o,v_{i,<t})}.
\label{eq:method_policy_ratio}
\end{equation}
This ratio is weighted by the group-relative advantage $A_i$, clipped within $[1-\epsilon_-,1+\epsilon_+]$, and regularized by a Kullback--Leibler (KL) divergence penalty with respect to the reference policy $\pi_{\mathrm{ref}}$. The policy maximizes
\begin{equation}
\begin{aligned}
\mathcal J_{\mathrm{GRPO}}(\theta)=\mathbb E\Bigg[\frac{1}{G}\sum_{i=1}^{G}\frac{1}{|v_i|}\sum_{t=1}^{|v_i|}\Big(&\min\{\rho_{i,t}A_i,\overline\rho_{i,t}A_i\}\\
&-\beta\mathcal K_{i,t}\Big)\Bigg],
\end{aligned}
\label{eq:method_grpo}
\end{equation}
where $\overline\rho_{i,t}=\operatorname{clip}(\rho_{i,t},1-\epsilon_-,1+\epsilon_+)$ and $\mathcal K_{i,t}$ denotes the KL regularization term relative to $\pi_{\mathrm{ref}}$. The clipping coefficients stabilize policy updates, while the KL coefficient $\beta$ controls deviation from the reference model. Following PPO-style clipped updates~\cite{schulman2017ppo}, this group-relative formulation avoids training a separate value model and is suitable for long-context RTL generation.

\subsection{Reward Model}
\label{subsec:method_reward_model}

\begin{figure}[!t]
\centering
\includegraphics[width=\columnwidth]{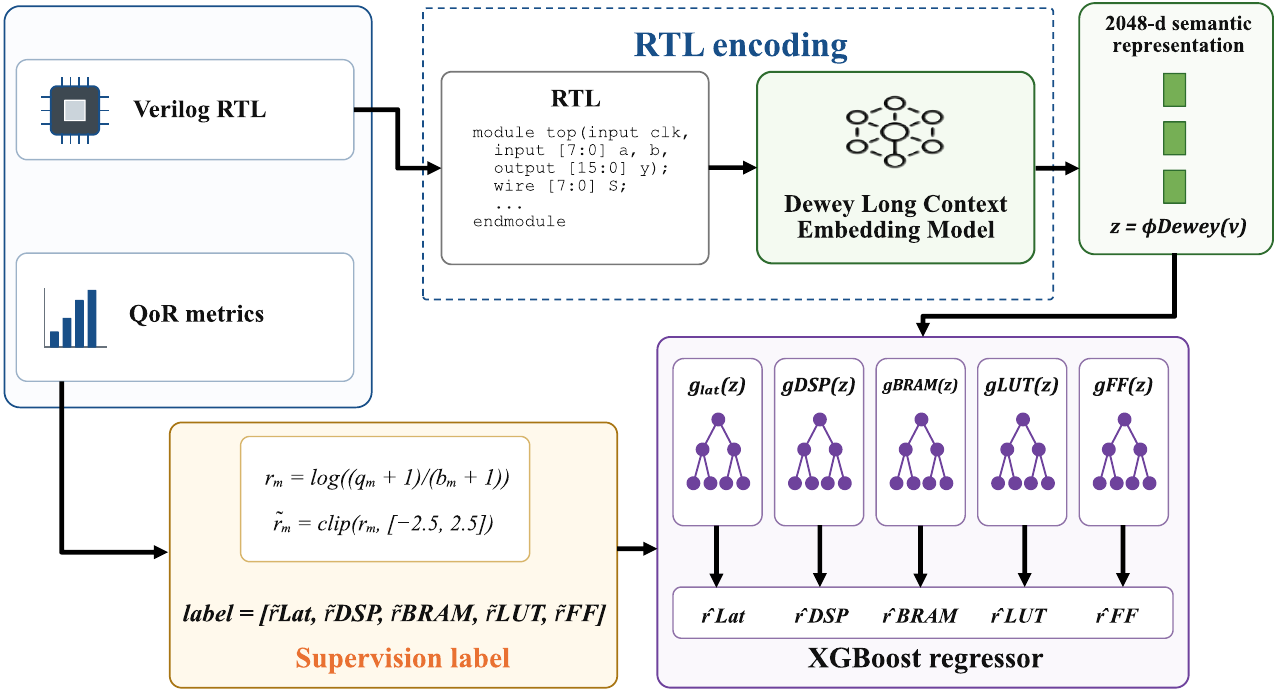}
\caption{Training pipeline of the QoR RM.}
\label{fig:method_reward_model}
\end{figure}

During GRPO, invoking real synthesis for every RTL generated in rollout would incur prohibitive computational cost. To address this issue, we design the QoR Prediction Reward Model (QoR RM), which approximates real-synthesis feedback with low-cost QoR prediction and converts it into continuous reward signals. The overall training pipeline of QoR RM is shown in Fig.~\ref{fig:method_reward_model}.

During QoR RM training, the training pipeline extracts the complete Verilog hierarchy from each synthesized design and encodes the RTL text with the Dewey long-context embedding model~\cite{zhang2025dewey}, yielding a 2048-dimensional representation:
\begin{equation}
\mathbf z=\phi_{\mathrm{Dewey}}(v)\in\mathbb R^{2048}.
\label{eq:method_embedding}
\end{equation}
Here, $v$ denotes the Verilog RTL, $\phi_{\mathrm{Dewey}}$ denotes the Dewey encoder, and $\mathbf z$ is the semantic representation of the RTL.

The supervision labels come from real-synthesis samples. For each sample, QoR RM predicts the five clipped candidate-to-baseline log ratios defined above:
\begin{equation}
\widetilde{\mathbf r}=[\widetilde r_{\mathrm{latency}},\widetilde r_{\mathrm{DSP}},\widetilde r_{\mathrm{BRAM}},\widetilde r_{\mathrm{LUT}},\widetilde r_{\mathrm{FF}}].
\label{eq:method_targets}
\end{equation}
Candidate-to-baseline log ratios provide region-normalized supervision, while the complete candidate RTL hierarchy jointly represents the underlying region computation and the realized microarchitecture. QoR RM therefore learns relative synthesis outcomes from RTL features that capture both design scale and candidate-specific optimization effects.

We train five single-target XGBoost regressors~\cite{chen2016xgboost}, one for each QoR target:
\begin{equation}
\widehat{\widetilde r}_m=g_m(\mathbf z),\qquad m\in\mathcal M.
\label{eq:method_predictor}
\end{equation}
Here, $g_m$ denotes the XGBoost regressor for target $m$.

During GRPO training, each candidate is first evaluated by the training-time functional validity gate described in the experimental protocol. Candidates that fail this functional check receive zero reward, whereas passing candidates are scored by QoR RM.

For candidates that pass the gate, QoR RM converts the predicted log ratios into latency and resource component utilities:
\begin{equation}
\begin{aligned}
\widehat U_{\mathrm{lat}}&=\operatorname{clip}_{[u_{\min},u_{\max}]}\bigl(\sigma(-\kappa\widehat{\widetilde r}_{\mathrm{latency}})\bigr),\\
\widehat{\widetilde r}_{\mathrm{res}}&=\sum_{m\in\mathcal R}w_m\widehat{\widetilde r}_m,\\
\widehat U_{\mathrm{res}}&=\operatorname{clip}_{[u_{\min},u_{\max}]}\bigl(\sigma(-\kappa\widehat{\widetilde r}_{\mathrm{res}})\bigr).
\end{aligned}
\label{eq:method_predicted_utilities}
\end{equation}
The reward is then composed according to the optimization objective in the prompt:
\begin{equation}
\begin{aligned}
\widehat S_{\mathrm{lat}}&=0.8\widehat U_{\mathrm{lat}}+0.2\widehat U_{\mathrm{res}},\\
\widehat S_{\mathrm{res}}&=0.2\widehat U_{\mathrm{lat}}+0.8\widehat U_{\mathrm{res}}.
\end{aligned}
\label{eq:method_predicted_scores}
\end{equation}
Let $F(x,v)\in\{0,1\}$ denote the outcome of the training-time functional check. The complete reward function is
\begin{equation}
R(x,o,v)=\begin{cases}
0,&F(x,v)=0,\\
\widehat S_{\mathrm{lat}}(v),&F(x,v)=1,\ o=\mathrm{latency},\\
\widehat S_{\mathrm{res}}(v),&F(x,v)=1,\ o=\mathrm{resource}.
\end{cases}
\label{eq:method_gated_reward}
\end{equation}
Thus, passing candidates generated from latency-oriented prompts are rewarded with $\widehat S_{\mathrm{lat}}$, while passing candidates generated from resource-oriented prompts are rewarded with $\widehat S_{\mathrm{res}}$.

\section{Experimental Setup}
\label{sec:experimental_setup}

\subsection{Implementation and Benchmark}
\label{subsec:implementation_benchmark}

All local models are trained on NVIDIA A800 GPUs with 80\,GB of memory using SWIFT and Megatron-SWIFT with low-rank adaptation (LoRA)~\cite{zhao2025swift,shoeybi2019megatron,hu2022lora}. The backbone is Qwen3-Coder-30B-A3B-Instruct~\cite{qwen2025coder30b}.

SFT is performed on six A800 GPUs with tensor, context, pipeline, and expert parallelism degrees of 2, 3, 1, and 2, respectively. The LoRA rank and alpha are 8 and 32, respectively. The learning rate is $1\times10^{-4}$, the warmup ratio is 0.05, and training takes approximately two days.

GRPO is performed on 32 A800 GPUs, with 16 allocated to
policy training and 16 to vLLM rollout~\cite{kwon2023vllm}.
The learning rate and KL coefficient are
$1.5\times10^{-6}$ and 0.001, respectively.
The lower and upper policy-ratio clipping coefficients
are $\epsilon_-=0.2$ and $\epsilon_+=0.28$.
The rollout group size is $G=4$, and the micro and global
batch sizes are 1 and 4, respectively, with the global
batch size counting candidate completions.
Each rollout batch comprises 16 completions from four
prompts, with four candidates per prompt, and is used
for four policy-update steps.
Training is configured for three epochs and takes
approximately 1.5 days. Inference is performed with vLLM on two A800 GPUs using a temperature of 0.2 and a top-$p$ of 0.9.

Five single-target XGBoost regressors for latency, DSP, BRAM, LUT, and FF are trained in parallel, each on a separate A800 GPU~\cite{chen2016xgboost}. The number and depth of trees, learning rate, subsampling ratios, and regularization hyperparameters are selected by grid search with ten-fold cross-validation. Training takes approximately 8.8 hours.

We construct an evaluation set from HLS-Eval, a benchmark for LLM-based HLS optimization~\cite{abikaram2025hlseval}. After excluding five cases without comparable output checks, we compare the remaining kernels against the original training kernels through source-provenance tracing and structural analysis. This process removes 30 overlapping cases, leaving 59 kernels. The retained kernels are provided as model inputs, while their C++ testbenches and expected outputs are reserved exclusively for final functional verification.

Final RTL simulation and synthesis are performed using Vivado 2023.2~\cite{amd2023vivado,amd2023simulation}, targeting the xcu280-fsvh2892-2L-e device with a 10\,ns clock-period constraint. The C++ outputs of the HLS baselines are compiled into RTL using Vitis HLS 2023.2~\cite{amd2023vitis} and then evaluated through the common hardware evaluation flow. All methods use the same device, clock constraint, and QoR scoring formula.

\subsection{Experimental Protocol}
\label{subsec:experimental_protocol}

For each kernel, we evaluate the base model, the SFT-only model, AutoC2V without the functional gate, and the complete AutoC2V model using two objective-conditioned prompts. Both prompts contain the same synthesizable C++ kernel and differ only in the optimization objective. The latency-oriented prompt prioritizes lower end-to-end latency, whereas the resource-oriented prompt prioritizes lower DSP, BRAM, LUT, and FF usage. We use the same prompts to query Claude Opus 4.8, GPT-5.5, Gemini 3.1 Pro Preview, DeepSeek v4 Pro, and GLM-5.2 through OpenRouter~\cite{openrouter2026quickstart}.

For comparison with HLSPilot and LLM-DSE, we reproduce their public workflows and connect their outputs to our synthesis and scoring flow~\cite{xiong2024hlspilot,wang2025llmdse}. The released LLM-DSE implementation supports Merlin-based compilation and QoR reporting but does not provide a directly executable Vitis HLS backend. We therefore replace its compilation and QoR-reporting interfaces with Vitis-compatible synthesis and report parsing. LLM-DSE's original multi-agent search logic is otherwise preserved, including its router, arbitrator, critic, and performance and resource specialists. We additionally include AutoDSE as a conventional HLS design-space exploration baseline~\cite{sohrabizadeh2022autodse}. We use an implementation adapted to the current Merlin and Vitis HLS environment, configure a 60-minute search limit for each kernel, and select the top-ranked output of the search workflow as the final design. AutoDSE's C++ output is compiled into RTL using Vitis HLS and then evaluated through the same final simulation, Vivado synthesis, and scoring flow as the other methods.

HLSPilot, LLM-DSE, and AutoDSE each retain one final design per kernel; their latency-oriented and resource-oriented scores are computed from the same implementation. AutoC2V instead generates separate outputs for the two objectives.

For models that directly generate RTL, functional correctness is evaluated through a C++-to-RTL co-simulation proxy used in two settings: the training-time functional validity gate and held-out test-set functional verification. The proxy exposes the same C++ signature as the target top function, allowing an existing C++ testbench to invoke the candidate RTL in place of the original C++ implementation. For each call, the proxy maps scalar and array arguments to RTL ports and memory or stream transactions. It then resets and starts RTL simulation, services data transfers, and waits for completion. Finally, the proxy writes the RTL outputs back to the corresponding C++ variables, allowing the testbench to execute its existing output checks.

During GRPO training of the complete AutoC2V model, the functional validity gate uses dedicated training testbenches constructed for the 201 source regions. GPT-5.5 assists in generating deterministic test cases and output-checking logic. Each region is assigned a fixed test suite containing boundary cases, such as zeros and signed extrema. The number of test cases depends on the region interface and array dimensions but remains fixed across all candidate RTL designs derived from that region. For each test case, the expected outputs are produced by executing the original region-level C++ implementation.

Before GRPO training of the complete AutoC2V model, each dedicated training testbench is verified to compile, terminate, and produce identical outputs across repeated executions. Candidate RTL designs are then evaluated through the proxy using the fixed test cases and output-checking logic. An unparseable module hierarchy or top-level interface is rejected before simulation. Candidates that fail interface parsing, compilation, simulation, timeout checks, completion detection, or output comparison receive zero reward without invoking QoR RM.

For held-out test-set functional verification, the HLS-Eval C++ testbenches are linked to the same proxy. Testbenches with built-in self-checking retain their input generation, expected data, numerical tolerances, and pass/fail logic. For testbenches without built-in pass/fail checks but with comparable output artifacts, we execute the original C++ implementation and the RTL proxy under identical input/output conditions and compare their observable outputs, including generated data or files. Thus, only the implementation invoked at the top-function boundary is replaced, while the benchmark's input generation and output-checking behavior remain unchanged.

All RTL designs, including those generated directly and those produced by Vitis HLS from the optimized C++ outputs of HLSPilot, LLM-DSE, and AutoDSE, undergo out-of-context (OOC) synthesis with Vivado 2023.2~\cite{amd2023vivado}. LUT, FF, DSP, and BRAM usage are obtained from the resulting utilization reports.

We define BRAM as total block-RAM usage and report it in 18-Kb equivalent units. Using the RAMB18 and RAMB36/FIFO categories in the Vivado utilization reports, we compute
\begin{equation}
B_{18K}=N_{\mathrm{RAMB18}}+2N_{\mathrm{RAMB36/FIFO}}.
\label{eq:setup_bram18k}
\end{equation}
Each RAMB36/FIFO unit is counted as two 18-Kb units.

Latency is measured through RTL simulation using Vivado Simulator~\cite{amd2023simulation}, with inputs supplied by the original HLS-Eval testbenches. We add an independent cycle counter to the testbench and, after reset deassertion, sample the DUT control signals 1\,ns after each rising clock edge. For the first transaction, we record the cycle $n_{\mathrm{start}}$ at which \texttt{ap\_start} is first observed high and the first cycle $n_{\mathrm{done}}$ at or after $n_{\mathrm{start}}$ at which \texttt{ap\_done} is observed high. Latency is computed as
\begin{equation}
\begin{aligned}
T_{\mathrm{latency}}
&=\left(n_{\mathrm{done}}-n_{\mathrm{start}}\right)T_{\mathrm{clk}},\\
T_{\mathrm{clk}}&=10\,\mathrm{ns}.
\end{aligned}
\label{eq:setup_latency}
\end{equation}

We additionally inspect the post-synthesis timing reports
for all designs included in the QoR statistics. All
implementations with a 10\,ns clock constraint satisfy
the applied timing constraints, with positive worst
negative slack (WNS) and zero total negative slack (TNS).
For the complete AutoC2V model, the minimum WNS values
are 1.777\,ns and 1.623\,ns under the latency and resource
objectives, respectively. Clockless combinational
implementations are excluded from clock-constraint statistics.

\section{Experimental Results and Analysis}
\label{sec:experimental_results}

This section evaluates AutoC2V in terms of functional correctness and hardware QoR through comparisons with commercial models and HLS optimization methods. Ablation and objective-weight sensitivity analyses examine the effects of the training strategy and different optimization preferences.

\subsection{Functional Correctness and Overall QoR}
\label{subsec:results_overall}

\begin{table*}[!t]
\caption{Functional correctness and QoR of direct RTL generation models.}
\label{tab:results_models}
\centering
\begingroup
\footnotesize
\setlength{\tabcolsep}{3pt}
\renewcommand{\arraystretch}{1.15}
\begin{tabular*}{\textwidth}{@{\extracolsep{\fill}}l*{8}{r}@{}}
\toprule
Method
& \multicolumn{2}{c}{Validity}
& \multicolumn{3}{c}{Latency-oriented prompt}
& \multicolumn{3}{c}{Resource-oriented prompt} \\
\cmidrule(lr){2-3}
\cmidrule(lr){4-6}
\cmidrule(l){7-9}
& Syn@1 & Func@1
& $U_{\mathrm{lat}}$ & $U_{\mathrm{res}}$ & Score
& $U_{\mathrm{lat}}$ & $U_{\mathrm{res}}$ & Score \\
\midrule
Qwen3-Coder-30B-A3B-Instruct
& 6.8\% & 0.0\%
& N/A & N/A & N/A
& N/A & N/A & N/A \\
SFT-only
& 70.3\% & 22.0\%
& 0.5357 & 0.4064 & 0.5098
& 0.2875 & 0.4928 & 0.4517 \\
AutoC2V w/o func.\ gate
& 76.3\% & 27.1\%
& 0.7002 & \textbf{0.4299} & 0.6462
& \textbf{0.3002} & 0.5999 & 0.5400 \\
AutoC2V
& \textbf{80.5\%} & \textbf{66.9\%}
& \textbf{0.8355} & 0.3789 & \textbf{0.7442}
& 0.2880 & \textbf{0.6747} & \textbf{0.5973} \\
\midrule
Claude Opus 4.8
& 70.3\% & 30.5\%
& 0.8123 & 0.2920 & 0.7083
& 0.1414 & 0.6428 & 0.5425 \\
GPT-5.5
& 56.8\% & 23.7\%
& 0.7733 & 0.3199 & 0.6826
& 0.0668 & 0.6422 & 0.5271 \\
Gemini 3.1 Pro Preview
& 49.2\% & 22.9\%
& 0.7406 & 0.3191 & 0.6563
& 0.0444 & 0.6395 & 0.5205 \\
DeepSeek v4 Pro
& 40.7\% & 19.5\%
& 0.6586 & 0.3220 & 0.5913
& 0.0903 & 0.6445 & 0.5337 \\
GLM-5.2
& 64.4\% & 22.0\%
& 0.7542 & 0.3618 & 0.6757
& 0.0945 & 0.6375 & 0.5289 \\
\bottomrule
\end{tabular*}
\par\smallskip
\begin{minipage}{\textwidth}
\footnotesize
\textit{Note:} Syn@1 and Func@1 are averaged over the two
optimization objectives. For each objective, utilities and
scores are arithmetic means over single-candidate outputs that
pass functional verification. Higher values are better.
N/A indicates no functionally passing output; w/o denotes
without. Bold values indicate column maxima.
\end{minipage}
\endgroup
\end{table*}

Table~\ref{tab:results_models} compares the base model, different AutoC2V training configurations, and five commercial models. Syn@1 and Func@1 denote the synthesis and functional pass rates of a single generated candidate, respectively, averaged over the two optimization objectives. QoR statistics are computed only from each model's RTL outputs that pass functional verification: latency utility, resource utility, and the objective-specific score are calculated for each implementation using the method in Section~\ref{sec:methodology}, and then averaged. Candidates that fail functional verification are excluded from these QoR means.

AutoC2V achieves a Syn@1 of 80.5\% and a Func@1 of 66.9\%, the highest among the compared models. Relative to SFT-only, these represent improvements of 10.2 and 44.9 percentage points, respectively. These results show that, after SFT establishes initial RTL generation capabilities, policy optimization combining functional gating and QoR rewards further improves generation quality, particularly functional correctness.

AutoC2V also achieves a higher functional pass rate than the commercial models. Claude Opus 4.8 obtains a Func@1 of 30.5\%, which AutoC2V exceeds by 36.4 percentage points. Thus, AutoC2V generates functionally passing RTL for more test kernels.

Among functionally passing outputs, AutoC2V achieves latency-oriented and resource-oriented scores of 0.7442 and 0.5973, respectively, both higher than those of the other models. Its latency-oriented outputs attain a latency utility of 0.8355, while its resource-oriented outputs attain a resource utility of 0.6747, demonstrating its ability to adapt hardware implementations to the requested objective. The base model produces no functionally passing output, so its QoR entries are marked N/A.

To compare implementation quality on the same kernels, we further perform per-kernel paired comparisons on kernels for which both AutoC2V and the corresponding commercial model pass functional verification. The horizontal axis shows the AutoC2V score minus the commercial-model score; positive values favor AutoC2V. Each point represents one common kernel, diamonds denote mean differences, and horizontal bars show percentile 95\% bootstrap intervals obtained from 10,000 resamples of the common kernels.

\begin{figure*}[!t]
\centering
\includegraphics[width=\textwidth]
{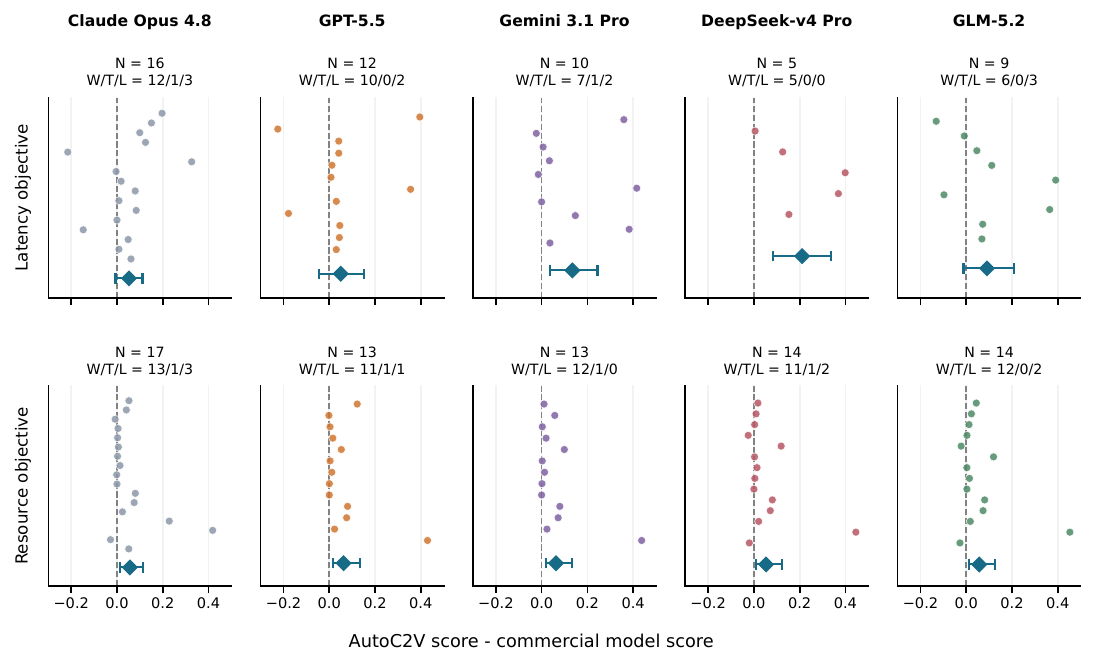}
\caption{Per-kernel paired QoR score differences between
AutoC2V and five commercial models. Columns correspond to
Claude Opus 4.8, GPT-5.5, Gemini 3.1 Pro, DeepSeek v4 Pro,
and GLM-5.2. The top and bottom rows show latency-oriented
and resource-oriented outputs, respectively. $N$ is the number
of common kernels, and W/T/L gives the numbers of wins, ties,
and losses for AutoC2V. Positive differences favor AutoC2V.
Diamonds and horizontal bars show mean differences and
percentile 95\% bootstrap intervals from 10,000 kernel resamples.}
\label{fig:results_commercial_paired}
\end{figure*}

For the latency objective, AutoC2V achieves win/tie/loss counts of 12/1/3, 10/0/2, and 7/1/2 against Claude Opus 4.8, GPT-5.5, and Gemini 3.1 Pro, with mean score differences of 0.0526, 0.0501, and 0.1345, respectively. Against DeepSeek v4 Pro, AutoC2V achieves a higher score on all five common kernels; against GLM-5.2, it wins on six of nine common kernels and loses on three. The corresponding mean differences are 0.2104 and 0.0908. All five comparisons yield positive mean differences, with AutoC2V achieving higher scores on most common kernels, as shown in Fig.~\ref{fig:results_commercial_paired}.

For the resource objective, the win/tie/loss counts against the five models are 13/1/3, 11/1/1, 12/1/0, 11/1/2, and 12/0/2, respectively. The corresponding mean differences are 0.0567, 0.0626, 0.0631, 0.0530, and 0.0569. All five bootstrap intervals lie above zero, showing consistent positive differences for this objective.

Together with Table~\ref{tab:results_models}, these results show that AutoC2V improves the functional pass rate and achieves higher mean objective-specific scores on common kernels for which both methods pass functional verification.

\subsection{Comparison with HLS Optimization Methods}
\label{subsec:results_hls}

\begin{table*}[!t]
\caption{QoR comparison with HLS optimization methods on common kernels.}
\label{tab:results_hls_common}
\centering
\begingroup
\small
\setlength{\tabcolsep}{4pt}
\renewcommand{\arraystretch}{1.15}
\begin{tabular*}{\textwidth}{@{\extracolsep{\fill}}l*{6}{r}@{}}
\toprule
Method
& \multicolumn{3}{c}{Latency objective}
& \multicolumn{3}{c}{Resource objective} \\
\cmidrule(lr){2-4}
\cmidrule(l){5-7}
& $U_{\mathrm{lat}}$ & $U_{\mathrm{res}}$ & Score
& $U_{\mathrm{lat}}$ & $U_{\mathrm{res}}$ & Score \\
\midrule
AutoC2V
& \textbf{0.7822} & 0.4087 & \textbf{0.7075}
& 0.2973 & \textbf{0.6794} & \textbf{0.6030} \\
AutoDSE
& 0.3616 & 0.1090 & 0.3111
& 0.3012 & 0.1121 & 0.1499 \\
HLSPilot
& 0.6664 & 0.3436 & 0.6019
& \textbf{0.6343} & 0.3767 & 0.4282 \\
LLM-DSE
& 0.5917 & \textbf{0.4871} & 0.5708
& 0.5939 & 0.5009 & 0.5195 \\
\bottomrule
\end{tabular*}
\par\smallskip
\begin{minipage}{\textwidth}
\footnotesize
\textit{Note:} All four methods are evaluated on the same
19 kernels for the latency objective and the same 24 kernels
for the resource objective. Higher utilities and scores are
better. Bold values indicate column maxima.
\end{minipage}
\endgroup
\end{table*}

Table~\ref{tab:results_hls_common} compares AutoC2V, AutoDSE, HLSPilot, and LLM-DSE on the same kernels. The latency and resource objectives include 19 and 24 common kernels, respectively.

AutoC2V achieves the highest mean objective-specific score for both objectives. Its latency-oriented score is 0.7075, exceeding those of AutoDSE, HLSPilot, and LLM-DSE by approximately 0.3964, 0.1057, and 0.1368, respectively. Its resource-oriented score is 0.6030, with corresponding improvements of approximately 0.4531, 0.1747, and 0.0835.

The component utilities further reveal the optimization preferences of the different methods. AutoC2V achieves a latency utility of 0.7822 under the latency objective and a resource utility of 0.6794 under the resource objective, both higher than those of the three HLS methods. Although some HLS methods achieve better secondary-objective utilities, AutoC2V performs better on the requested primary objective and the combined score.

\begin{figure*}[!t]
\centering
\includegraphics[width=\textwidth]
{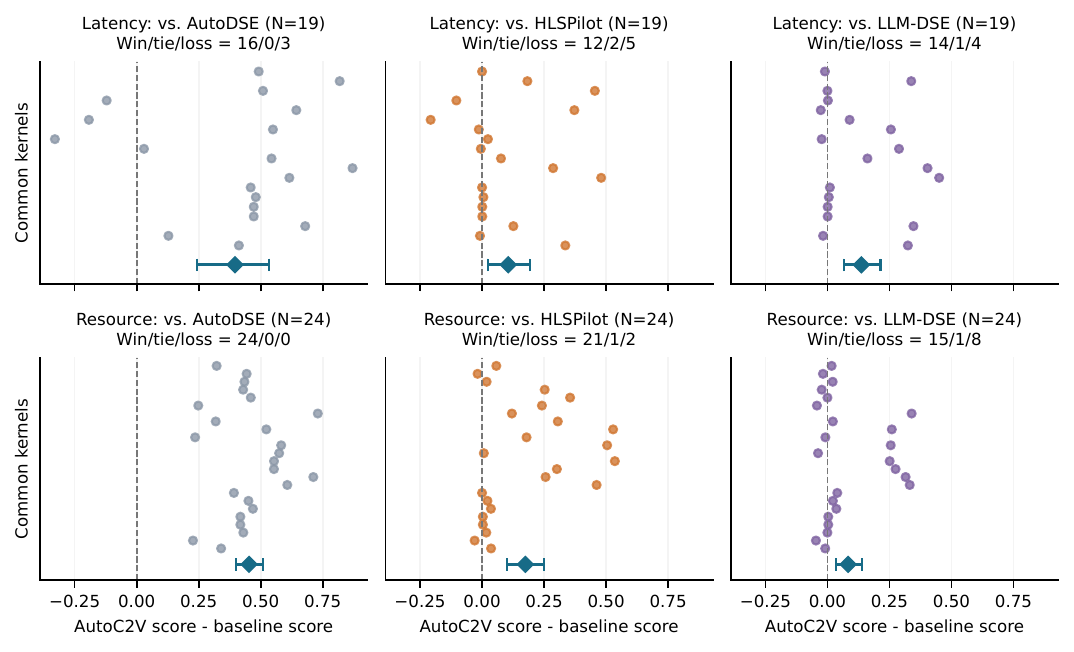}
\caption{Per-kernel paired QoR score differences between
AutoC2V and HLS optimization methods on common kernels.
Columns compare AutoC2V with AutoDSE, HLSPilot, and LLM-DSE.
The top row uses the 19 common kernels for the latency
objective, and the bottom row uses the 24 common kernels for
the resource objective. Each point represents one kernel;
positive differences favor AutoC2V. Diamonds and horizontal
bars show mean differences and percentile 95\% bootstrap
intervals from 10,000 kernel resamples. Win/tie/loss counts
are reported in each panel.}
\label{fig:results_hls_paired}
\end{figure*}

Fig.~\ref{fig:results_hls_paired} shows the per-kernel results using the same score-difference definition and interval-estimation procedure as the commercial-model comparisons. For the latency objective, AutoC2V achieves win/tie/loss counts of 16/0/3, 12/2/5, and 14/1/4 against AutoDSE, HLSPilot, and LLM-DSE, respectively. For the resource objective, the corresponding counts are 24/0/0, 21/1/2, and 15/1/8.

All six comparisons yield positive mean differences, with the displayed bootstrap intervals lying above zero. These per-kernel results show that the mean advantage is accompanied by score improvements on most common kernels. The comparison methods nevertheless remain competitive on some kernels. For example, LLM-DSE achieves higher scores on eight common kernels under the resource objective, reflecting differences in the suitability of optimization methods for individual kernels.

\subsection{Ablation of the Training Strategy and Functional Gate}
\label{subsec:results_ablation}

\begin{table*}[!t]
\caption{Ablation of the training strategy and functional gate.}
\label{tab:results_ablation}
\centering
\begingroup
\small
\setlength{\tabcolsep}{3pt}
\renewcommand{\arraystretch}{1.15}
\begin{tabular*}{\textwidth}{@{\extracolsep{\fill}}l*{8}{r}@{}}
\toprule
Configuration
& \multicolumn{2}{c}{Validity}
& \multicolumn{3}{c}{Latency-oriented prompt}
& \multicolumn{3}{c}{Resource-oriented prompt} \\
\cmidrule(lr){2-3}
\cmidrule(lr){4-6}
\cmidrule(l){7-9}
& Syn@1 & Func@1
& $U_{\mathrm{lat}}$ & $U_{\mathrm{res}}$ & Score
& $U_{\mathrm{lat}}$ & $U_{\mathrm{res}}$ & Score \\
\midrule
SFT-only
& 70.3\% & 22.0\%
& 0.5357 & 0.4064 & 0.5098
& 0.2875 & 0.4928 & 0.4517 \\
AutoC2V w/o func.\ gate
& 76.3\% & 27.1\%
& 0.7002 & \textbf{0.4299} & 0.6462
& \textbf{0.3002} & 0.5999 & 0.5400 \\
AutoC2V
& \textbf{80.5\%} & \textbf{66.9\%}
& \textbf{0.8355} & 0.3789 & \textbf{0.7442}
& 0.2880 & \textbf{0.6747} & \textbf{0.5973} \\
\bottomrule
\end{tabular*}
\par\smallskip
\begin{minipage}{\textwidth}
\footnotesize
\textit{Note:} SFT-only omits GRPO; w/o func.\ gate denotes
GRPO without the training-time functional gate. Syn@1 and
Func@1 are averaged over the two objectives. QoR means are
computed only from each configuration's outputs that pass
final functional verification. Higher values are better.
Bold values indicate column maxima.
\end{minipage}
\endgroup
\end{table*}

Table~\ref{tab:results_ablation} compares SFT-only, AutoC2V without the functional gate, and the complete AutoC2V model. As in Table~\ref{tab:results_models}, QoR means are computed only from the RTL outputs that pass functional verification for each configuration.

Compared with SFT-only, GRPO without the functional gate achieves higher scores for both objectives. For the latency objective, latency utility increases from 0.5357 to 0.7002 and the score increases from 0.5098 to 0.6462. For the resource objective, resource utility increases from 0.4928 to 0.5999 and the score increases from 0.4517 to 0.5400. These results show that, under the evaluated training configuration, QoR-driven policy optimization improves the objective-specific quality of functionally passing RTL outputs.

The complete AutoC2V model further increases latency utility and score under the latency objective to 0.8355 and 0.7442, and resource utility and score under the resource objective to 0.6747 and 0.5973. Meanwhile, Func@1 increases from 27.1\% without the gate to 66.9\%. These results show that combining functional gating with QoR rewards improves the functional pass rate while achieving better objective-specific QoR.

Changes in the component utilities also reflect the priority of the requested objective. Under the latency objective, the complete model achieves the highest latency utility, while resource utility decreases. Under the resource objective, the improvement is concentrated in resource utility, with relatively little change in latency utility. This behavior is consistent with the reward design, which prioritizes the requested primary objective while accounting for the secondary objective.

\subsection{Validation of QoR Reward Model Predictions}
\label{subsec:results_qor_rm}

To evaluate the predictive performance of QoR RM, we test
it on the previously reported RTL outputs of the complete
AutoC2V model that pass functional verification.
Table~\ref{tab:results_qor_rm} reports the mean squared error
(MSE), mean absolute error (MAE), coefficient of determination
($R^2$), and Pearson correlation coefficient for latency
and the four resource metrics.

\begin{table}[!t]
\caption{Predictive performance of QoR RM.}
\label{tab:results_qor_rm}
\centering
\begingroup
\footnotesize
\setlength{\tabcolsep}{3pt}
\renewcommand{\arraystretch}{1.15}
\begin{tabular*}{\columnwidth}
{@{\extracolsep{\fill}}lrrrr@{}}
\toprule
Target & MSE & MAE & $R^2$ & Pearson \\
\midrule
Latency & 0.0411 & 0.1325 & 0.9850 & 0.9925 \\
DSP     & 0.0166 & 0.1020 & 0.9869 & 0.9937 \\
BRAM    & 0.0005 & 0.0168 & 0.9933 & 0.9967 \\
LUT     & 0.0081 & 0.0711 & 0.9945 & 0.9973 \\
FF      & 0.0268 & 0.1140 & 0.9845 & 0.9926 \\
\bottomrule
\end{tabular*}
\par\smallskip
\begin{minipage}{\columnwidth}
\footnotesize
\textit{Note:} Lower MSE and MAE indicate smaller prediction
errors; higher $R^2$ and Pearson correlation coefficients
indicate better fit and stronger linear correlation,
respectively.
\end{minipage}
\endgroup
\end{table}

Across the evaluated RTL outputs, all five prediction
targets achieve $R^2$ values of at least 0.9845 and Pearson
correlation coefficients of at least 0.9925, indicating
close agreement between the predictions and measured targets.
For latency, the $R^2$ and Pearson correlation coefficient
are 0.9850 and 0.9925, respectively. For the four resource
metrics, $R^2$ ranges from 0.9845 to 0.9945, while Pearson
correlation coefficients range from 0.9926 to 0.9973.

These results provide empirical support for the predictive
performance of QoR RM on AutoC2V-generated RTL and its use
as an approximate QoR evaluator during policy optimization.

\subsection{Sensitivity to Objective Weights}
\label{subsec:results_weight_sensitivity}

\begin{figure*}[!t]
\centering
\includegraphics[width=\textwidth]
{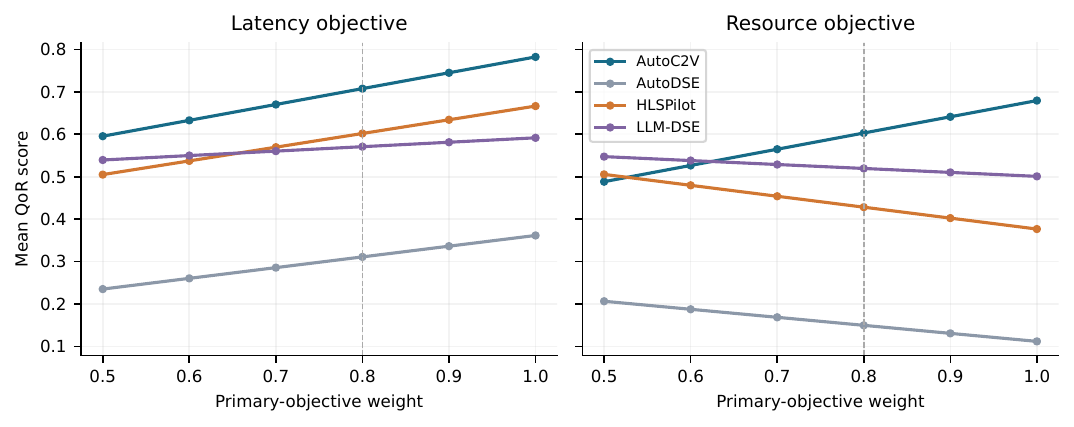}
\caption{Sensitivity of mean objective-specific scores to
the primary-objective weight $\alpha$. The left and right
panels show latency-oriented and resource-oriented outputs,
respectively. Outputs and common-kernel sets are held fixed
while scores are recomputed using different weights. The
default resource-type weights are retained. The dashed
vertical line marks the default primary-objective weight,
$\alpha=0.8$.}
\label{fig:results_weight_sensitivity}
\end{figure*}

To examine the effect of evaluation preferences, we hold each method's outputs and the common-kernel sets fixed, vary the primary-objective weight, and recompute the scores. Let $\alpha$ denote the primary-objective weight:
\begin{equation}
\begin{aligned}
S_{\mathrm{lat}}(\alpha)
&=\alpha U_{\mathrm{lat}}+(1-\alpha)U_{\mathrm{res}},\\
S_{\mathrm{res}}(\alpha)
&=(1-\alpha)U_{\mathrm{lat}}+\alpha U_{\mathrm{res}}.
\end{aligned}
\label{eq:results_weight_sensitivity}
\end{equation}
Fig.~\ref{fig:results_weight_sensitivity} considers six values of $\alpha$ from 0.5 to 1.0, with the dashed line indicating the default setting of $\alpha=0.8$.

For the latency objective, AutoC2V achieves the highest mean score at every evaluated weight, maintaining its advantage across different latency preferences. Its score increases as more weight is assigned to latency, consistent with its higher latency utility.

For the resource objective, the ranking depends more strongly on the evaluation preference. LLM-DSE achieves a higher combined score when latency and resources receive approximately equal weights. AutoC2V achieves the highest score at all evaluated resource weights of 0.7 or greater. This agrees with the component utilities in Table~\ref{tab:results_hls_common}: AutoC2V's resource-oriented outputs achieve higher resource utility, whereas some HLS methods achieve higher latency utility. Consequently, AutoC2V's advantage becomes more pronounced as resource optimization receives greater emphasis.

\section{Discussion and Limitations}
\label{sec:discussion_limitations}

The experimental results show that learning the relationship
between hardware implementations and QoR from real synthesis
data can improve direct C++-to-RTL generation. AutoC2V
generates different implementations according to the
optimization objective and jointly uses functional gating
and QoR rewards to guide policy optimization. The
objective-weight sensitivity analysis further shows that
relative performance depends on evaluation preferences:
AutoC2V maintains higher scores across the evaluated latency
weights, while its advantage under the resource objective
becomes more pronounced as the resource weight increases.
In practical applications, optimization preferences should
therefore be selected according to latency requirements and
resource budgets, and generated implementations should be
assessed using the individual component metrics.

QoR RM reduces the cost of repeatedly invoking synthesis
tools during policy optimization. The evaluation in
Section~\ref{subsec:results_qor_rm} demonstrates close
agreement between predicted and measured targets for RTL
QoR, supporting its use as an approximate QoR evaluator.
Its reliability across a broader range of RTL structures
remains to be examined.

The main generation evaluation uses 59 held-out kernels
under a common FPGA device and clock constraint. Latency
is measured through RTL simulation, and resource utilization
is obtained from out-of-context synthesis. The results
therefore primarily characterize generation capabilities
and optimization effects within this evaluation environment.
Larger kernels, more complex interfaces, other target
devices, and post-place-and-route timing and resource
utilization remain to be evaluated.

\section{Conclusion}
\label{sec:conclusion}

We presented AutoC2V, a framework driven by real synthesis
feedback for generating QoR-optimized Verilog RTL directly
from synthesizable C++ kernels. By combining real synthesis
data, QoR RM guidance, a training-time functional validity
gate, and GRPO, AutoC2V substantially improves functional
pass rates over the unfine-tuned and SFT-only models.
It also achieves higher mean objective-specific QoR scores
than representative HLS approaches (HLSPilot, LLM-DSE,
and AutoDSE) and advanced commercial LLMs on common kernels.
Future work will address more complex control logic,
multi-kernel designs, and cross-FPGA generation.

\section*{Acknowledgment}
During data construction, Qwen3-Next-80B-A3B-Thinking assisted in supplying missing code elements, and GPT-5.5 assisted in source-region decomposition, as described in Section~\ref{subsec:method_data}. GPT-5.5 also assisted in generating deterministic training test cases and output-checking logic, as described in Section~\ref{subsec:experimental_protocol}. The authors are responsible for the final technical content, experimental results, and references.

\bibliographystyle{IEEEtran}
\bibliography{references}

\end{document}